\documentclass[twocolumn,showpacs,preprintnumbers,amsmath,amssymb]{revtex4}

\usepackage{graphicx}
\usepackage{dcolumn}
\usepackage{bm}
\usepackage{booktabs}
\begin{document}
\title{Influence of uniaxial tensile stress on the mechanical and piezoelectric properties of short-period ferroelectric superlattice}

\author{Yifeng Duan$^1$}%
\email{yifeng@semi.ac.cn}
\author{Chunmei Wang$^2$}
\author{Gang Tang$^1$}%
\author{Changqing Chen$^{3}$}%
\email{chencq@tsinghua.edu.cn}

\affiliation{$^1$Department of Physics, China University of Mining
and Technology, Xuzhou 221116, People's Republic of
China\\$^2$School of Aerospace, Xi'an Jiaotong University, Xi'an,
710049, People's Republic of China\\$^3$Department of Engineering
Mechanics, AML, Tsinghua University, Beijing 100084, People's
Republic of China}

\date{\today}

\begin{abstract}
Tetragonal ferroelectric/ferroelectric $\rm{BaTiO_3/PbTiO_3}$
superlattice under uniaxial tensile stress along the $c$ axis is
investigated from first principles. We show that the calculated
ideal tensile strength is 6.85 GPa and that the superlattice under
the loading of uniaxial tensile stress becomes soft along the
nonpolar axes. We also find that the appropriately applied uniaxial
tensile stress can significantly enhance the piezoelectricity for
the superlattice, with piezoelectric coefficient $d_{33}$ increasing
from the ground state value by a factor of about 8, reaching 678.42
$\rm{pC/N}$. The underlying mechanism for the enhancement of
piezoelectricity is discussed.
\end{abstract}

\maketitle

\section{INTRODUCTION}
Ferroelectrics which can convert mechanical to electrical energy
(and vice versa) have wide applications in medical imaging,
telecommunication and ultrasonic devices, the physical properties of
which are sensitive to external conditions, such as strain, film
thickness, temperature, electric and magnetic fields
\cite{r1,r2,r3}. ${\rm BaTiO_3}$ (BTO) and ${\rm PbTiO_3}$ (PTO), as
prototype ferroelectric materials and simple systems, have been
intensively studied \cite{r4,r5}. It is known that the
ferroelectricity arises from the competition of short-range
repulsions which favor the paraelectric cubic phase and Coulomb
forces which favor the ferroelectric phase \cite{r6,r7}. As the
pressure increases, the short-range repulsions increase faster than
the Coulomb forces, leading to the reduced ferroelectricity.
Accompanied with the suppression of ferroelectricity, the
piezoelectricity decreases and even disappears. However, recent
studies have shown that the noncollinear polarization rotation,
occurring at phase transition pressure, can result in the giant
piezoelectric response \cite{r8,r9}. In contrast to previous
theoretical studies of the effects of epitaxial strain on the
spontaneous polarization of ferroelectric thin films, we have
systematically studied the influence of uniaxial and in-plane
epitaxial strains on the mechanical and piezoelectric properties of
perovskite ferroelectrics \cite{r101,r102,r10,r11,r12,r13}. So far,
there has been no previous work on the effect of uniaxial tensile
strains on the mechanical and piezoelectric properties of
short-period BTO/PTO superlattices.

Ferroelectric superlattices composed of alternating epitaxial oxides
ultrathin layers are currently under intensive study due to their
excellent ferroelectric and piezoelectric properties \cite{r14}.
Ferroelectricity can be induced in $\rm{AB_1O_3/AB_2O_3}$
superlattice in spite of the paraelectric nature of $\rm{AB_1O_3}$
and $\rm{AB_2O_3}$. This is because the coincidence of the positive
and negative charge centers is destroyed in the superlattice and
electric dipoles are induced. Moreover, ferroelectricity can be
enhanced in ferroelectric superlattices in certain stacking
sequences \cite{r15}. The overall polarization of three-component
$\rm{SrTiO_3(STO)/BTO/PTO}$ ferroelectric superlattices can also be
improved by increasing the number of BTO and PTO layers \cite{r16}.
Thanks to the periodic nature, it is possible to study the effect of
uniaxial or biaxial strains on the properties of ferroelectric
superlattices from first principles.

In this work, we perform total energy as well as linear response
calculations to study the effect of uniaxial tensile stress along
the $c$ axis on the mechanical and piezoelectric properties of
short-period BTO/PTO superlattice. We show the mechanical properties
by calculating the ideal tensile strength, elastic constants and
valence charge density at different strains. We also show the
influence of uniaxial stress on the piezoelectricity. To reveal the
underlying mechanisms, we study the effects of uniaxial tensile
stress on the atomic displacements and Born effective charges,
respectively.

\section{COMPUTATIONAL METHODS}
Our calculations are performed within the local density
approximation (LDA) to the density functional theory (DFT) as
implemented in the plane-wave pseudopotential ABINIT package
\cite{r17}. To ensure good numerical convergence, the plane-wave
energy cutoff is set to be 80 Ry and the Brillouin zone integration
is performed with $6\times6\times6$ {\bf{k}}-meshpoints. The
norm-conserving pseudopotentials generated by the OPIUM program are
tested against the all-electron full-potential linearized augmented
plane-wave method \cite{r18,r19}. The orbitals of Ba ${\rm
5s^25p^66s^2}$, Pb ${\rm 5d^{10}6s^26p^2}$, Ti ${\rm
3s^23p^63d^24s^2}$, and O ${\rm 2s^22p^4}$ are explicitly included
as valence electrons. The dynamical matrices and Born effective
charges are computed by the linear response theory of strain type
perturbations, which has been proved to be highly reliable for
ground state properties \cite{r20,r21,r22}. The polarization is
calculated by the Berry-phase approach \cite{r23}. The LDA is used
instead of the generalized gradient approximation (GGA) because the
GGA is found to overestimate both the equilibrium volume and strain
for the perovskite structures \cite{r24}. The piezoelectric strain
coefficients $d_{i{\nu}}$=$\Sigma_{\mu=1}^{6}e_{i\mu}s_{{\mu}\nu}$,
where $\textbf{e}$ is the piezoelectric stress tensor and the
elastic compliance tensor $\textbf{s}$ is the reciprocal of the
elastic stiffness tensor $\textbf{c}$ (Roman indexes from 1 to 3,
and Greek ones from 1 to 6).

In the calculations, a double-perovskite ten-atom supercell along
the $c$ axis is used for the tetragonal short-period BTO/PTO
superlattice. The primitive periodicity of tetragonal structure with
the space group $P4mm$ is retained, which is more stable in energy
than the rhombohedral structure. For the tetragonal perovskite
structure compounds BTO and PTO, the equilibrium lattice parameters
are $a$(BTO)=3.915 $\rm{\AA}$, $c$(BTO)=3.995 $\rm{\AA}$,
$a$(PTO)=3.843 $\rm{\AA}$, and $c$(PTO)=4.053 $\rm{\AA}$, which are
slightly less than the experimental values of 3.994, 4.034, 3.904,
and 4.135 $\rm{\AA}$, respectively \cite{r11,r12}. A sketch of
ground-state short-period BTO/PTO superlattice with its atomic
positions is shown in Fig.1.

\begin{figure}[here]
\includegraphics[width=8.5cm,height=7.5cm]{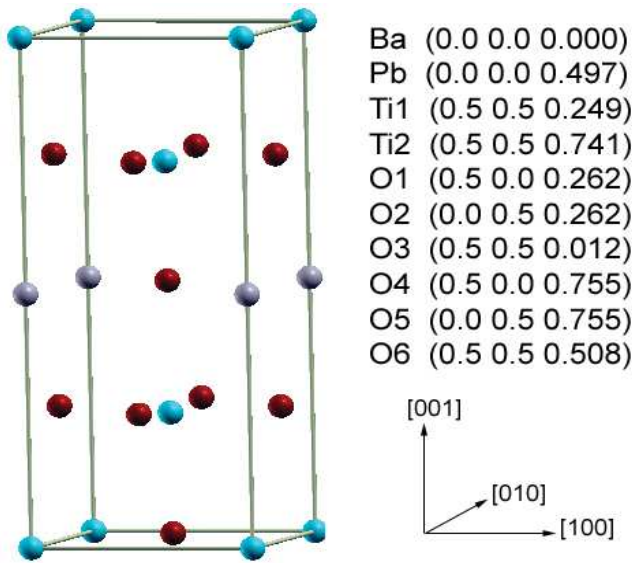}
\caption{The sketch of short-period ferroelectric superlattice with
its atomic positions.}
\end{figure}

To calculate the uniaxial tensile stress $\sigma_{33}$, we apply a
small strain increment $\eta_3$ along the $c$ axis and then conduct
structural optimization for the lattice vectors perpendicular to the
$c$ axis and all the internal atomic positions until the two
components of stress tensor (i.e., $\sigma_{11}$ and $\sigma_{22}$)
are smaller than 0.05 GPa. The strain is then increased step by
step. Since $\sigma_{11}$=$\eta_1(c_{11}+c_{12})+\eta_3c_{13}$, the
elastic constants satisfy
$\eta_3$/$\eta_1$$\approx$-$(c_{11}+c_{12})$/$c_{13}$ under the
loading of uniaxial tensile strain applied along the $c$ axis, where
the strains $\eta_{i}$ are calculated by
$\eta_{1}=\eta_{2}=(a-a_0)/a_0$ and $\eta_{3}=(c-c_0)/c_0$, with
$a_0$=3.897 $\rm{\AA}$ and $c_0$=7.859 $\rm{\AA}$ being the lattice
constants of the unstrained superlattice structure. We have examined
the accuracy of our calculations by studying the influence of
different strains on the properties of BTO and PTO, respectively
\cite{r10,r11,r12,r13}.

\section{RESULTS AND DISCUSSION}
\begin{figure}[here]
\includegraphics[width=8.5cm]{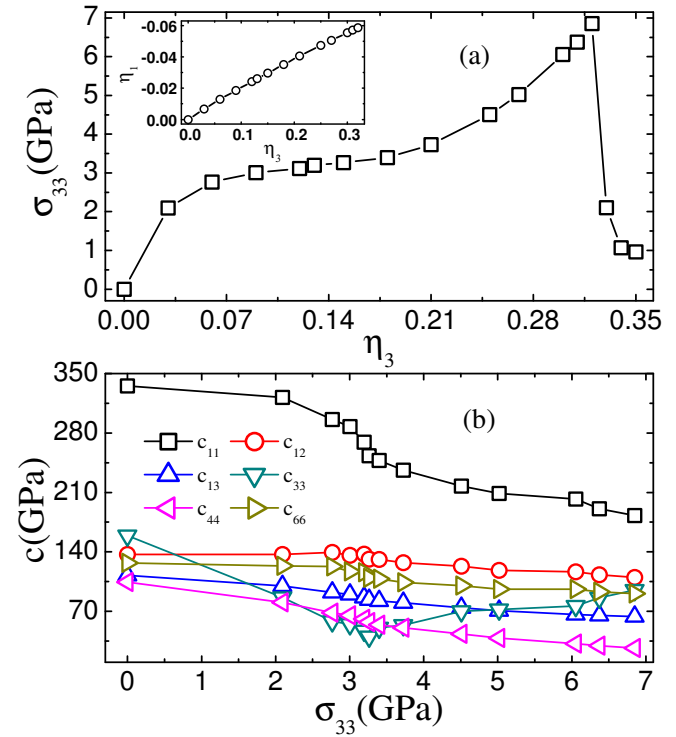}
\caption{(a) Uniaxial tensile stress as a function of tensile strain
$\eta_3$, and the inset reflects the relation between strains
$\eta_3$ and $\eta_1$. (b) Elastic constants as a function of stress
$\sigma_{33}$.}
\end{figure}

Figure 2(a) shows the uniaxial tensile stress $\sigma_{33}$ as a
function of strain $\eta_3$. The relation between strains $\eta_1$
and $\eta_3$ is shown in the inset, which satisfy $\eta_3>-2\eta_1$.
The stress $\sigma_{33}$ increases until reaching its maximum value
of 6.85 GPa with increasing strain, indicating that the calculated
ideal tensile strength is 6.85 GPa for the superlattice, which is
the maximum stress required to break the superlattice. Figure 2(b)
shows the elastic constants as a function of stress $\sigma_{33}$,
which reflect the relation between stress and strain. What is the
most unexpected is that the constant $c_{33}$ first decreases until
reaching its minimum value at $\sigma_c$=3.26 GPa and then gradually
increases, promising a large electromechanical response at
$\sigma_c$ \cite{r25}. The minimum $c_{33}$ corresponds to the
minimum slope of the curve of Fig.2(a) at $\sigma_c$. Other elastic
constants, especially $c_{11}$, always decrease with increasing
$\sigma_{33}$, indicating that the superlattice under the loading of
uniaxial tensile stress along the $c$ axis becomes soft along the
nonpolar axes.

\begin{figure}[here]
\includegraphics[width=8.7cm]{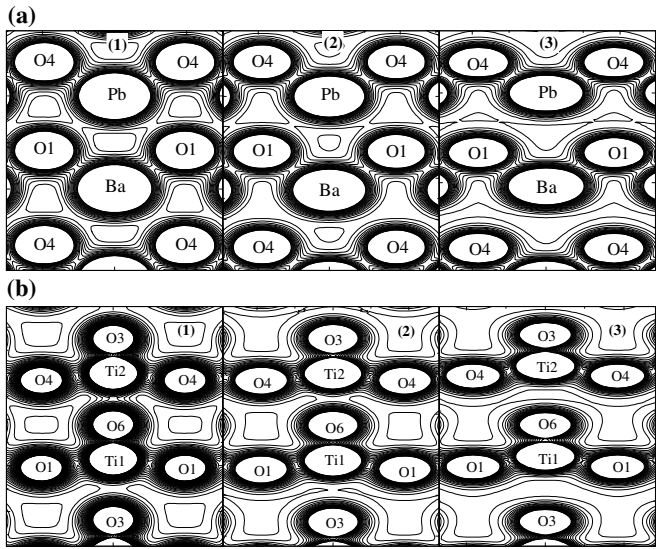}
\caption{Calculated valence charge density along the $c$ axis in the
$(100)$ (a) and $(200)$ (b) planes of superlattice at equilibrium
(1), maximum piezoelectric coefficient (2) and ideal tensile
strength (3).}
\end{figure}

To illustrate the change of chemical bonds with uniaxial tensile
stress, figures 3(a) and 3(b) are plotted to show the valence charge
density along the $c$ axis in the $(100)$ and $(200)$ planes of the
superlattice at equilibrium, maximum piezoelectric coefficient and
ideal tensile strength, respectively. The Pb-$\rm{O_4}$,
Ba-$\rm{O_1}$, $\rm{Ti_1}$-$\rm{O_1}$ and $\rm{Ti_2}$-$\rm{O_4}$
bond lengths are not sensitive to the uniaxial tensile stress along
the $c$ axis, suggesting that the orbital hybridizations between
these atoms are not sensitive to the uniaxial strain, whereas the
$\rm{Ti_1}$-$\rm{O_3}$ and $\rm{Ti_2}$-$\rm{O_6}$ bonds elongate
remarkably with increasing stress. Following the evolution of the
charge density, we find that the weak $\rm{Ti_1}$-$\rm{O_3}$ bond
starts to break first, followed by the $\rm{Ti_2}$-$\rm{O_6}$ bond.
After the bond breaks, the system converts into a planar structure
with alternating layers. On the other hand, figures 3(a) and 3(b)
show that the valence charge density becomes more and more
unsymmetrical with the uniaxial tensile stress increasing,
indicating the increase in polarization. To confirm this, we have
directly calculated the relations between the polarization and the
uniaxial tensile stress with the Berry-phase approach.

\begin{figure}[here]
\includegraphics[width=8.5cm]{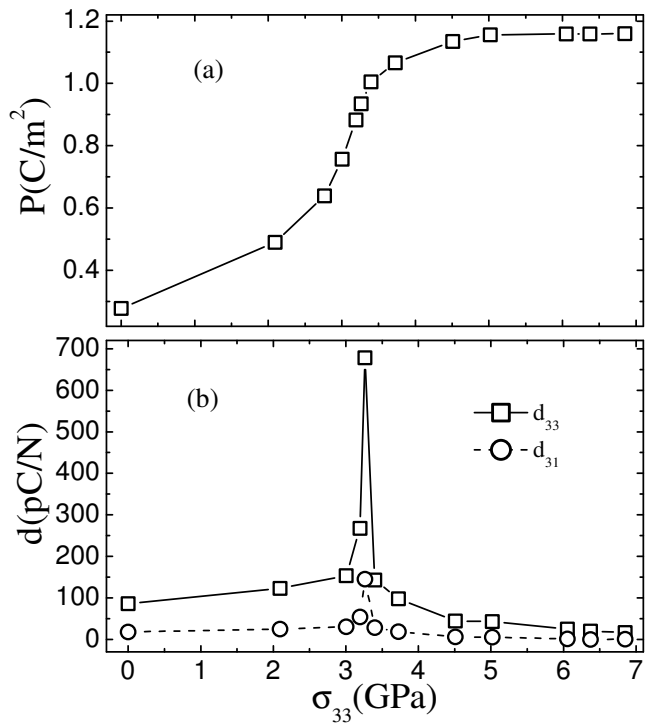}
\caption{Uniaxial tensile stress dependence of (a) polarization and
(b) piezoelectric coefficients (i.e., $d_{31}$ and $d_{33}$).}
\end{figure}

Figure 4(a) shows the polarization as a function of uniaxial tensile
stress. For the ground-state superlattice, the calculated
spontaneous polarization of 0.29 $\rm{C/m^2}$ is less than the
theoretical value of 0.81 $\rm{C/m^2}$ of ground-state PTO, but
slightly larger than the value of 0.28 $\rm{C/m^2}$ of tetragonal
BTO (the other theoretical value is 0.26 $\rm{C/m^2}$ \cite{r26}),
which supports the conclusion that the sharp interfaces suppress the
polarization in short-period BTO/PTO superlattices \cite{r26}. As
the stress $\sigma_{33}$ increases, the polarization dramatically
increases with the maximum slope appearing at $\sigma_c$, indicating
that the ferroelectric phase becomes more and more stable with
respect to the paraelectric phase.

Figure 4(b) shows the variation of piezoelectric coefficients with
stress $\sigma_{33}$, which are calculated by the linear response
theory. The piezoelectric coefficients all increase with increasing
$\sigma_{33}$ and reach their maximum values at $\sigma_c$,
indicating that the appropriately applied uniaxial tensile stress
can enhance the piezoelectricity for the superlattice. The
piezoelectric coefficient $d_{33}$ of ground-state superlattice is
86.36 $\rm{pC/N}$, which is slightly less than the value of 103.18
$\rm{pC/N}$ of PTO, but much larger than the value of 36.43
$\rm{pC/N}$ of BTO. Under the loading of uniaxial tensile stress
applied along the $c$ axis, $d_{33}$ is increased from its ground
state value by a factor of about 8, reaching 678.42 $\rm{pC/N}$ for
the superlattice. From previous calculations \cite{r12}, we know
that the uniaxial tensile stress can only enhance $d_{33}$ of PTO to
the maximum value of 380.50 $\rm{pC/N}$. The enhancement of
piezoelectricity is supported by the conclusion of uniaxial tensile
stress dependency of elastic constant $c_{33}$ [see Fig.2(b)]. Note
that the polarization under uniaxial stress remains along the
$<001>$ direction and that the piezoelectric coefficients reflect
the slope of polarization versus stress curves. The enhancement of
piezoelectricity corresponds to the maximum slope of the curve of
Fig.4(a) at $\sigma_c$, it is the change of magnitude of
polarization that leads to the enhancement of piezoelectricity.

\begin{figure}[here]
\includegraphics[width=8.5cm]{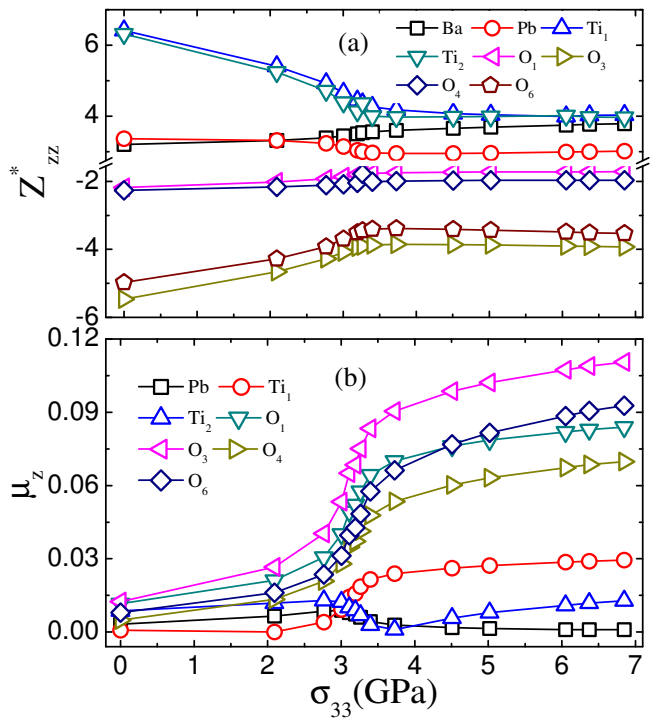}
\caption{(a) Born effective charges $\rm{Z^*_{zz}}$ and (b) atomic
displacements along the $c$ axis (in $c$ units), relative to the
centrosymmetric reference structure, as a function of uniaxial
tensile stress.}
\end{figure}

To reveal the underlying mechanisms for the abnormal
piezoelectricity, we study the effects of uniaxial tensile stress on
the Born effective charges and atomic displacements, respectively
[see Figs.5(a) and 5(b)]. Since the atomic displacements and
polarization are all along the $c$ axis, only charges
$\rm{Z^*_{zz}}$ contribute to the polarization. The uniaxial tensile
stress reduces the effective charges, which remain almost constant
when $\sigma_{33}>\sigma_c$. The charges $\rm{Z^*_{zz}}$ of
$\rm{O_1}$ and $\rm{O_4}$ atoms are much close to their normal
charges, so does the case of Ti atoms when $\sigma_{33}>\sigma_c$,
whereas $\rm{Z^*_{zz}}$ of $\rm{O_3}$ and $\rm{O_6}$ atoms are
anomalously large compared with their normal charges, suggesting the
strong orbital hybridization between $\rm{Ti_1}$ (and $\rm{Ti_2}$)
3$d$ and $\rm{O_6}$ (and $\rm{O_3}$) 2$p$ states [see Fig.3(b)].
Note that the Ba atom is fixed at (0, 0, 0) during the
first-principles simulations. The displacements of O atoms, which
are much larger than those of Pb and Ti atoms for a broad range of
stress, are greatly enhanced as the stress $\sigma_{33}$ increases,
especially near $\sigma_c$, leading to the drastic increase in
polarization. It is concluded that as the stress $\sigma_{33}$
increases, the atomic displacements are so greatly enhanced that the
overall effect is the increase in polarization, even though the
magnitudes of $\rm{Z^*_{zz}}$ decrease with the stress increasing.

\section{SUMMARY}
In summary, we have studied the influence of uniaxial tensile stress
applied along the $c$ axis on the mechanical and piezoelectric
properties of short-period BTO/PTO superlattice using
first-principles methods. We show that the calculated ideal tensile
strength is 6.850 GPa and that the superlattice under the loading of
uniaxial tensile stress becomes soft along the nonpolar axes. We
also find that the appropriately applied uniaxial tensile stress can
significantly enhance the piezoelectricity for the superlattice. Our
calculated results reveal that it is the drastic increase of atomic
displacements along the $c$ axis that leads to the increase in
polarization and that the enhancement of piezoelectricity is
attributed to the change in the magnitude of polarization with the
stress. Our work suggests a way of enhancing the piezoelectric
properties of the superlattices, which would be helpful to enhance
the performance of the piezoelectric devices.

\begin{acknowledgments}
The work is supported by the National Natural Science Foundation of
China under Grant Nos. 10425210, 10832002 and 10674177, the National
Basic Research Program of China (Grant No. 2006CB601202), and the
Foundation of China University of Mining and Technology.
\end{acknowledgments}

\clearpage

\end{document}